\documentclass[11pt]{article}
\usepackage{graphicx, setspace, amssymb, subfigure, url, multirow, booktabs, verbatim, bm,threeparttable,algorithm, color} 
\usepackage{natbib}
\usepackage{graphicx} 
\usepackage{booktabs,color}
\usepackage{booktabs,float,multirow}
\usepackage{booktabs,color,epsfig,graphicx}  
\usepackage{amsmath,amssymb}
\usepackage{iftex}
\usepackage{makecell}
\usepackage{booktabs,float,multirow}
 
\newtheorem{theorem}{Theorem}

\allowdisplaybreaks[3]
\usepackage{hyperref}	
\hypersetup{colorlinks,%				% Reference color setup	
	linkcolor=blue,%
	citecolor=blue}

\begin{document}

	\def\spacingset#1{\renewcommand{\baselinestretch}%
		{#1}\small\normalsize} \spacingset{1}

	%%%%%%%%%%%%%%%%%%%%%%%%%%%%%%%%%%%%%%%%%%%%%%%%%%%%%%%%%%%%%%%%%%%%%%%%%%%%%%
	
	\setlength{\textheight}{575pt}
\setlength{\baselineskip}{23pt}
\def\spacingset#1{\renewcommand{\baselinestretch}%
	{#1}\small\normalsize} \spacingset{1}
\title{Supervised Bayesian Joint Graphical Model for Simultaneous Network Estimation and Subgroup Identification}
\author{Xing Qin$^{1}$, Xu Liu$^2$,
	Shuangge Ma$^{3}$, and Mengyun Wu$^{2,*}$\\ \\
	$^{1}$School of Statistics and Data Science, Shanghai University of International Business and Economics \\
	$^{2}$School of Statistics and Data Science, Shanghai University of Finance and Economics \\
	$^{3}$Department of Biostatistics, Yale School of Public Health\\
	\\
	email: wu.mengyun@mail.shufe.edu.cn}
\date{} 
\maketitle
	
	\bigskip
	\begin{abstract} 
Heterogeneity is a fundamental characteristic of cancer. To accommodate heterogeneity, subgroup identification has been extensively studied and broadly categorized into unsupervised and supervised analysis. Compared to unsupervised analysis, supervised approaches potentially hold greater clinical implications. Under the unsupervised analysis framework, several methods focusing on network-based subgroup identification have been developed, offering more comprehensive insights than those restricted to mean, variance, and other simplistic distributions by incorporating the interconnections among variables. However, research on supervised network-based subgroup identification remains limited. In this study, we develop a novel supervised Bayesian graphical model {(SBJGM)} for jointly identifying multiple heterogeneous networks and subgroups. In the proposed model, heterogeneity is not only reflected in molecular data but also associated with a clinical outcome, and a novel similarity prior is introduced to effectively accommodate similarities among the networks of different subgroups, significantly facilitating clinically meaningful biological network construction and subgroup identification. The consistency properties of the estimates are rigorously established, and an efficient algorithm is developed. Extensive simulation studies and a real-world application to {The Cancer Genome Atlas (TCGA)} data are conducted, which demonstrate the advantages of the proposed approach in terms of both subgroup and network identification. 
\end{abstract}

\noindent%
{\it Keywords:}  Bayesian analysis; Finite mixture of regression models; Gaussian graphical model; Heterogeneous network estimation.%3 to 6 keywords, that do not appear in the title
\vfill

\newpage
\spacingset{1.8} % DON'T change the spacing!

\section{Introduction} 
 \label{s:intro}

 For cancer, heterogeneity is a fundamental characteristic and is essential for unraveling the inherent complexities of biological systems. To describe and accommodate heterogeneity, researchers have conducted extensive studies on subgroup identification with high-dimensional molecular data, which can be roughly classified into { ``unsupervised” and ``supervised”. 
In this paper, the terms ``unsupervised” and ``supervised” are defined based on the type of information used. ``Unsupervised” means that the method relies solely on the distributions of molecular variables, whereas ``supervised” means that the method leverages predictor-outcome associations to inform heterogeneity modeling.
} More specifically, unsupervised approaches typically emphasize differences in the means of molecular factors {$\boldsymbol{X}$} (Table \ref{fig:intro}, a), with examples including \cite{Pan2007PenalizedMC} and \cite{2018Sparse}, and variances (Table \ref{fig:intro}, b), as exemplified by \cite{Xie2008PenalizedMC}. Under the supervised analysis framework, a popular approach is the finite mixture of regression models, which captures the heterogeneity of genetic effects on clinical outcome {$Z$} (Table \ref{fig:intro}, d). Examples include both Bayesian approaches \citep{Raman2010InfiniteMM,ahmad2017towards} and penalization approaches \citep{2019AOAS,Zhong2023}. These models enable the integration of information from both clinical outcomes and molecular data to identify clinically relevant subgroups. Despite significant achievements, existing techniques remain inadequate as they focus on the distributional properties or the influences of individual variables but neglect the interconnections among variables. There is a strong need for more informative approaches for subgroup identification.  
 
\begin{table}[ht]
\centering
\footnotesize
\caption{{Comparison of unsupervised and supervised subgroup identification models, where $\boldsymbol{X}$ denotes the molecular variables and $Z$ denotes the clinical outcome.}}
\label{fig:intro}
\begingroup
\renewcommand{\arraystretch}{0.6}
\begin{tabular}{@{}p{3.7cm}p{12cm}@{}}
\toprule
\textbf{Type} & \textbf{Model and Example} \\
\midrule
\multirow{3}{*} {\makecell[l]{\textbf{Unsupervised}\\ (relying solely on  the \\ heterogeneous  \\  distributions   of  $\boldsymbol{X}$)}}
& \textbf{a. Mean-based}:  \( \boldsymbol{X} \sim \mathcal{N}(\boldsymbol{\mu}_k, \boldsymbol{V}) \), where $\boldsymbol{\mu}_k$ is the mean vector in the $k$th subgroup and $\boldsymbol{V}$ is the diagonal covariance matrix.
Example: \( \boldsymbol{\mu}_1 = (1,0,0)^\top, \ \boldsymbol{\mu}_2 = (0,1,-1)^\top \). \\

& \textbf{b. Variance-based}: \( \boldsymbol{X} \sim \mathcal{N}(\boldsymbol{\mu}_k, \boldsymbol{D}_k) \), where $\boldsymbol{D}_k$ is the diagonal covariance matrix in the $k$th subgroup.  Example: \( \boldsymbol{D}_1 = \text{diag}(1,2,1), \ \boldsymbol{D}_2 = \text{diag}(2,1,2) \). \\

& \textbf{c. Network-based}: \( \boldsymbol{X} \sim \mathcal{N}(\boldsymbol{\mu}_k, \boldsymbol{\Omega}_k^{-1}) \), where $\boldsymbol{\Omega}_k$ is the precision matrix in the $k$th subgroup.  
Example: \( \boldsymbol{\Omega}_1 = \begin{pmatrix} 1 & -1 & 0 \\ -1 & 2 & 0 \\ 0 & 0 & 1 \end{pmatrix}, \ \boldsymbol{\Omega}_2 = \begin{pmatrix} 2 & 0 & -1 \\ 0 & 1 & 0 \\ -1 & 0 & 2 \end{pmatrix} \). \\
\midrule
  {\textbf{Supervised}  (leveraging heterogeneous associations  
  between $Z$ and $\boldsymbol{X}$)}
& \textbf{d. Linear effect-based}: \( Z \sim \mathcal{N}(\boldsymbol{\beta}_k^\top \boldsymbol{X}, \sigma_k^2) \), where $\boldsymbol{\beta}_k$ and $\sigma_k$ are the regression parameter vector and standard deviation in the $k$th subgroup.
Example: \( \boldsymbol{\beta}_1 = (-1,0,0)^\top, \ \boldsymbol{\beta}_2 = (0,1,-1)^\top; \ \sigma_1 = 1, \ \sigma_2 = 2 \). \\
\bottomrule
\end{tabular}
\endgroup
\end{table}

 The importance of network-based subgroup identification has been emphasized in recent literature \citep{ren2022gaussian}. Subtype-specific networks are pivotal in understanding complex biological processes. They utilize information embedded in the interconnections among variables (Table \ref{fig:intro}, c), providing richer insights than marginal means and variances. As illustrated in the literature \citep{danaher2014joint,peterson2015bayesian}, in addition to heterogeneity, there can be commonalities across subgroup-specific networks, which can help improve estimation efficiency and enhance interpretability through information borrowing. % \citep{2016Laurent} yang2015fused,

 Among the diverse statistical methods for network analysis, Gaussian graphical models (GGMs) have been prominent. Here, the nodes of a network are represented by random variables with a joint Gaussian distribution, while the edges reflect conditional dependencies between nodes that are determined by non-zero entries of the inverse covariance matrix (or precision matrix) \citep{gan2019bayesian,colombi2024learning}. To detect unknown heterogeneity, several network-based methods have been proposed under the framework of the Gaussian mixture model (GMM), with particular attention to identifying both structural differences and potential commonalities among multiple networks. For example, \cite{gao2016estimation} incorporated a joint fused graphical lasso penalty into GMM for learning the related structures across different unknown subgroups. Similarly, a {Simultaneous Clustering And estimatioN (SCAN)} method proposed by \cite{hao2018simultaneous} used a joint group graphical lasso penalty for simultaneous subgroup identification and joint graphical model estimation. \cite{Qin2024multinet} proposed decomposing the GGM into a set of sparse regression problems and introduced a composite minimax concave penalty for jointly estimating multiple networks. Under the Bayesian framework, \cite{dai2022bayesian} proposed a {Bayesian Clustering and multiple Graphical Selection (BCGS)} method that imposes a joint spike-and-slab graphical lasso prior on the precision matrices.  

 However, these GMM-based approaches belong to the unsupervised paradigm, focusing solely on the discovery of molecularly separable subgroups, do not consider any clinical outcomes, and may lead to subgroups lacking sufficient clinical implications. Published literature has demonstrated that supervised heterogeneity analysis that incorporates clinical information is promising in identifying clinically relevant subgroups and provides a more comprehensive understanding of disease \citep{2019AOAS,Zhong2023}.
 However, the existing supervised heterogeneity methods predominantly focus on detecting heterogeneous predictor-outcome associations without appropriately accommodating the network structures among predictors. Overall, there is a notable gap in jointly exploring molecular network structures and conducting subgroup identification in a supervised manner.

 In this paper, we develop a novel supervised network-based heterogeneity analysis approach to simultaneously conduct network estimation and subgroup identification. This approach is among the first to conduct joint estimation for multiple networks in a supervised manner with unknown subgroup structures. Significantly advancing from the existing unsupervised network-based subgroup identification approaches \citep{hao2018simultaneous,li2019bayesian,dai2022bayesian}, we develop a novel Bayesian framework integrating a finite mixture of Gaussian graphical models with supervised regression models to accommodate heterogeneity involving not only the distributions of predictors but also the predictor-outcome associations, leading to clinically more  meaningful biological network construction and subgroup identification. Additionally, advancing from the existing joint network analysis based on the group penalties \citep{danaher2014joint,hao2018simultaneous,li2019bayesian,dai2022bayesian}, a novel similarity prior is introduced to capture the similarity information between the values of connection effects of distinct networks, facilitating more effective information borrowing and more appropriate modeling of the underlying biological mechanisms. 
 Furthermore, unlike the existing supervised subgroup identification approaches \citep{ahmad2017towards,Zhong2023}, we fully take the specific genetic networks into account while simultaneously accommodating the clinical information. The consistency properties of the proposed estimators are rigorously established, achieved through an efficient algorithm. Comprehensive simulation studies are conducted, demonstrating the advantages of the proposed approach in terms of subgroup and network identification. Additionally, a real-world application to TCGA data highlights the practical utilization of our approach.

 \section{Methods}\label{s:methods} 
 We consider data with a survival outcome. Modeling survival presents more challenges due to non-negativity and censoring, and similar approaches can be applied to other clinical outcomes. Let $T$, $\Delta$, and $\boldsymbol{X} = \left(X_1,\ldots,X_p\right)^\top$ represent the logarithm of the minimum of survival time and censoring time, censoring indicator, and $p$-dimensional vector of predictors (e.g., gene expressions), respectively. Here, $T=\operatorname{min}\left(Z,C\right)$ with $Z$ and $C$ being the logarithms of survival time and censoring time, and $\Delta={I}\{ Z\leq C\}$ with ${I}\{\cdot\}$ being the indicator function.  Assume there are $n$ subjects $\left\{\left(t_i,\delta_i,\boldsymbol{x}_i\right),i=1,\ldots,n\right\}$, which form $K$ subgroups. 
 Denote $\boldsymbol{z}, \boldsymbol{c}, \boldsymbol{\delta},$ and $\boldsymbol{t}$ as the vectors of $z_i$'s, $c_i$'s, $\delta_i$'s, and $t_i$'s, respectively, and $\textbf{X}$ as the $n\times p$ matrix composed of $\boldsymbol{x}_i$'s.
 \subsection{Bayesian mixture modeling} 
 Denote $g_i \in\{1, \ldots, K\}$ as the subgroup membership of the $i$th subject. Consider the accelerated failure time (AFT) model: 
 $$z_i =\ddot{\beta}_{0k}+\ddot{\boldsymbol{\beta}}_k^{\top}\boldsymbol{x}_{i}+\epsilon_i,	\text{~when~}g_i=k,$$
 where $\ddot{\beta}_{0k}$ and  $\ddot{\boldsymbol{\beta}}_k=\left(\ddot{\beta}_{k,1},\ldots,\ddot{\beta}_{k,p}\right)^{\top}$ are the subgroup-specific intercept and regression parameters, and $\epsilon_i$ is a normally distributed random variable with mean $0$ and precision parameter ${\tau}_k^2$.
 We propose the following Bayesian mixture model:
 {
 \begin{equation}\label{equ:model}
 	\begin{aligned} 
 		&  t_i\mid\left(\delta_i, z_i,c_i\right) =z_i^{\delta_i}  c_i^{1-\delta_i},
        \\&
        \delta_i \mid\left(z_i,c_i\right)  = {I}\left\{z_i\leq  c_i\right\}, 
        \\&
        c_i\sim f_C(c_i),
 		\\&	z_i \mid \left(\boldsymbol{\beta},\boldsymbol{x}_{i},  g_i=k;\boldsymbol{\tau}\right)\sim \mathcal{N}\left(z_i\mid\beta_{0k}/\tau_k+\boldsymbol{\beta}_k^{\top}\boldsymbol{x}_{i}/\tau_k, \tau_k^{-2}\right),
        \\&\boldsymbol{x}_{i} \mid\left(\boldsymbol{\mu},\boldsymbol{\Omega},g_i=k\right)	\sim  \mathcal{N}\left(\boldsymbol{x}_i\mid \boldsymbol{\mu}_{k}, \boldsymbol{\Omega}_{k}^{-1}\right).
 	\end{aligned}
 \end{equation}  
 }
Here,  $f_C$ is the density function of the logarithm of censoring time $C$, $\boldsymbol{\mu}_{k}=\left(\mu_{k,1},\cdots,\mu_{k,p}\right)^\top$ and $\boldsymbol{\Omega}_k=\left(\omega_{k,jl}\right)_{p\times p}$ are the mean and precision matrix for the predictors in the $k$th subgroup, and $\boldsymbol{\beta}$, $\boldsymbol{\tau}$, $\boldsymbol{\mu}$, and  $\boldsymbol{\Omega}$ are the vectors of $\left({\beta}_{0k},\boldsymbol{\beta}_k\right)$'s, ${\tau}_k$'s, $\boldsymbol{\mu}_k$'s, and $\boldsymbol{\Omega}_k$'s, respectively, with ${\boldsymbol{\beta}}_k=\ddot{\boldsymbol{\beta}}_k \tau_k$ and $\beta_{0k}=\ddot{\beta}_{0k} \tau_k$. For the $k$th subgroup, we estimate the network based on $\boldsymbol{\Omega}_k$, where $\omega_{k,jl}\neq 0$ corresponds to a connection between predictors $l$ and $j$.

 After integrating out  $\boldsymbol{z}$ and $\boldsymbol{c}$, we have the joint distribution of $\left(\boldsymbol{t},\boldsymbol{\delta},\textbf{X}\right)$ as:
 \begin{equation*}
 	\begin{aligned}
 		& f\left(\boldsymbol{t},\boldsymbol{\delta},\textbf{X}\mid \boldsymbol{\beta},\boldsymbol{\mu},\boldsymbol{\Omega} ,\boldsymbol{\tau},\boldsymbol{g}\right)=\prod_{i=1}^n \prod_{k=1}^K \left[f_k\left(t_i, \boldsymbol{x}_i,\delta_i \mid  {\boldsymbol{\mu}}_k, {\boldsymbol{\Omega}}_k, {\boldsymbol{\beta}}_{k}, {\beta}_{0k}, {{\tau}}_{k}\right)\right]^{I\left\{g_i=k\right\}}\\=&\prod_{i=1}^n \prod_{k=1}^K  \left\{
 		f_{\boldsymbol{X}}\left(\boldsymbol{x}_i\mid \boldsymbol{\mu}_{k}, \boldsymbol{\Omega}_{k}\right)\left[f_Z\left(t_{i} \mid  \boldsymbol{x}_i, \boldsymbol{\beta}_k,\beta_{0k},{\tau}_k\right)S_C(t_i)\right]^{\delta_i}\left[f_C(t_i)S_Z\left(t_{i} \mid  \boldsymbol{x}_i, \boldsymbol{\beta}_k,\beta_{0k},{\tau}_k\right)\right]^{1-\delta_i}\right\}^{{I}\left\{g_i=k\right\}} ,  
 	\end{aligned}
 \end{equation*}
 where $f_{\boldsymbol{X}}$ and $f_{Z}$ are the density functions of $\boldsymbol{X}$ and $Z$ given in (\ref{equ:model}), respectively, and $S_C$ and $S_Z$ are the survival functions of $C$ and $Z$, respectively.
 
 In (\ref{equ:model}),  we assume random censoring with $c_i$ being independent of $z_i$ and $\boldsymbol{x}_i$, as in the literature \citep{2019AOAS}. We propose combining the AFT model with the mixture Gaussian graphical model to accommodate the multi-layer heterogeneity. {The log-normal AFT model is adopted for simplicity, interpretability, and ease of exposition, which is common in current practice.} We introduce a subgroup-specific coefficient $\boldsymbol{\beta}_k$ to explicitly capture the heterogeneous associations of the molecular predictors with survival time. A reparameterization technique based on the noise level is adopted for $\boldsymbol{\beta}_k$ to  ensure that the proposed estimator remains scale-invariant under the mixture model \citep{stadler2010,Zhong2023}. This technique facilitates an ``adaptive adjustment'' of the coefficients, aligning them with data characteristics and leading to more accurate estimations. On the other hand, within the Gaussian graphical model, we utilize $\boldsymbol{\mu}_k$ and $\boldsymbol{\Omega}_k$ to characterize heterogeneity attributed to both differential mean expressions and differential network structures. Consequently, the network's heterogeneity is not only intricately linked to the interested measurements but also closely connected to the distribution of survival time, making the resulting networks clinically more meaningful. For example, researchers can better understand the differences between patients with good survival and those with bad survival on specific biological networks, thereby more effectively guiding subsequent treatments. Here, consistent with prior studies \citep{hao2018simultaneous, dong2024tuning}, we assume that the subgroup number $K$ is {known}. 
 
% Different from \cite{gan2019bayesian}, which studies a single homogeneous network, our work considers supervised heterogeneous multiple networks. Different from \cite{hao2018simultaneous}, which considers an unsupervised mixture model without outcome information, our method captures heterogeneity in both predictor distributions and predictor-outcome associations, providing a more comprehensive and clinically relevant characterization. 
 
  { Our methodological framework advances beyond existing unsupervised or single-network approaches, such as \cite{gan2019bayesian} and \cite{hao2018simultaneous}. While these prior methods do not incorporate clinical outcomes and are limited to either unsupervised mixture modeling or a single graphical model, our work introduces a supervised, multi-network approach. Methodologically, we move beyond frequentist regularization or single-graph Bayesian priors by developing a novel Bayesian hierarchical model. This model integrates censored time-to-event outcomes through an AFT model and employs a structured prior to jointly estimate multiple heterogeneous networks (Sections \ref{prior} and \ref{prior2}). By doing so, our approach captures heterogeneity in both predictor distributions and predictor-outcome associations, offering a more comprehensive and clinically interpretable characterization.
 }
 
 \subsection{Prior specification}\label{prior}
 We propose the following priors for the parameters in (\ref{equ:model}):
\begin{equation*}
 	  {\begin{aligned} 
 		&  
 		 {\omega_{k,jj}\sim \operatorname{Exp}(\omega_{k,jj};\tau_0)}, \text{ for } j=1,\ldots,p, 	
 		\\& 	\omega_{k, jl} \mid \gamma_{k, jl}  \sim \gamma_{k, jl} \operatorname{LP}\left(\omega_{k, jl} ; v_{1}\right)+\left(1-\gamma_{k, jl}\right) \operatorname{LP}\left(\omega_{k, jl} ; v_0\right),\text{ for } j<l, 
 		\\&
        \gamma_{k, jl}  \sim \operatorname{Bern}\left(p_1\right), \text{ for } j<l, 	  \\&
        \beta_{k,j}  \sim \operatorname{LP}\left(\beta_{k,j};1/\lambda_1\right)
 		,
        \\&\mu_{k,j} \sim \operatorname{LP}\left(\mu_{k,j};1/\lambda_2\right),	
        \\&g_{i} \mid \boldsymbol{\pi_{g}}\sim \operatorname{Multi}\left(\pi_{1}, \pi_{2}, \ldots, \pi_{K}\right).
 	\end{aligned}}
 \end{equation*}
 Here, $\operatorname{Exp}(\omega_{k,jj};\tau_0)$ is the Exponential distribution with parameter $\tau_0$, $\gamma_{k,jl}$ is a binary indicator corresponding to $\omega_{k, jl}$,  $\operatorname{LP}(y;v)$ is the Laplace distribution with a scale parameter $v$, $p_1\in(0,1)$ is the prior probability for the Bernoulli distribution,  and $\boldsymbol{\pi}_g=\left(\pi_1,\ldots,\pi_K\right)^{\top} $ is the vector composed of the Multinomial distribution's parameters, which specifies the prior probabilities that each sample is drawn from different subgroups. $\tau_0$, $\lambda_1$, $\lambda_2$, and $v_{1}>v_{0}$ are the positive tuning parameters.  
 
 In the proposed Bayesian framework, we first introduce a weakly informative Exponential prior to the diagonal elements $\omega_{k,jj}$'s of the precision matrices, which are not subject to shrinkage, following \cite{gan2019bayesian}. We introduce a spike-and-slab Laplace prior \citep{rovckova2018bayesian,yang2021gembag} for $\omega_{k,jl}$ based on the indicator $\gamma_{k, jl}$, for   
 identifying individual sparse patterns of the $K$ networks. With this prior, when $\gamma_{k, jl}=1$, the corresponding entry $\omega_{k, jl}$ comes from the slab component with a relative large variance $2v_1^2$ and is likely to take a value away from zero; when $\gamma_{k, jl}=0$, $\omega_{k, jl}$ comes from the spike component with a small variance $2v_0^2$ that can induce values towards zero. Compared to the Laplace prior, the spike-and-slab Laplace prior involves two different scale parameters $v_1$ and $v_0$ to distinguish active from ignorable coefficients and can nonlinearly adapt to coefficients with $\gamma_{k, jl}$, achieving ``selective shrinkage'' with a smaller bias. { The spike-and-slab Laplace prior utilizes latent indicators $\gamma_{k,jl}$ to facilitate a rigorous assessment of Bayesian uncertainty. Rather than relying on the point estimates of $\omega_{k,jl}$, this approach performs network construction via the posterior inclusion probabilities %(PIPs) 
 derived from $\gamma_{k,jl}$. This probabilistic framework effectively quantifies model-selection uncertainty while ensuring theoretical tractability and computational efficiency.} %This Laplace distribution-based prior can automatically achieve sparsity via the $L_1$ norm on $\omega_{k,jl}$. This advances from the commonly adopted spike-and-slab Gaussian prior which yields nonsparse posterior modes for $\omega_{k,jl}$'s and relies on a post-data thresholding-based selection strategy utilizing $\gamma_{k, jl}$. The strategy that is based on the estimation and sparsity of the continuous $\omega_{k,jl}$ can lead to more effective identification of sparse posterior patterns \citep{rovckova2018bayesian} and can also facilitate a simpler theoretical study and a more efficient computation. 

 { Additionally, to accommodate the sparse structure of  truly associated predictors and predictor means, sparse priors are assumed for ${\beta}_{k,j}$ and $\mu_{k,j}$ for regularized estimation. Since only a subset of molecular means contributes to subgroup differentiation, imposing sparsity on 
$\mu_{k,j}$ is essential in high-dimensional clustering, and estimators without such regularization can be sub-optimal \citep{hao2018simultaneous}. Given that estimating ${\beta}_{k,j}$ and $\mu_{k,j}$ is generally less demanding than estimating the precision matrix entries $\omega_{k,jl}$, we employ a Laplace prior for both sets of parameters. This choice provides a suitable trade-off between effective regularization and computational efficiency.
}
  
 \subsection{Additional priors for accommodating common structures}\label{prior2}
 To borrow information from cross-subgroup similarity, we further introduce a latent vector {$\boldsymbol{\theta}_{jl}=\left(\theta_{1,jl},\ldots,\theta_{K,jl}\right)^\top$ for $\boldsymbol{\omega}_{jl}=\left(\omega_{1,jl},\ldots,\omega_{K,jl}\right)^\top$} and assume a Gaussian distribution with the precision matrix based on a similarity-based matrix $\boldsymbol{L}^{(jl)}$,
\[\boldsymbol{\theta}_{jl}\sim\mathcal{N}\left(\boldsymbol{0},\left(u\boldsymbol{L}^{(jl)}\right)^{-1}\right), \quad
 \boldsymbol{\theta}_{jl}\mid \boldsymbol{\omega}_{jl} \sim {I}\left\{	\boldsymbol{\theta}_{jl}=	\boldsymbol{\omega}_{jl}\right\},
\]
 with $u$ being the positive tuning parameter. Here, $\boldsymbol{\theta}_{jl}\sim\mathcal{N}\left(\boldsymbol{0},\left(u\boldsymbol{L}^{(jl)}\right)^{-1}\right)$ can be interpreted as a generative model in which a zero vector is observed from a Gaussian distribution parameterized by $\boldsymbol{\theta}_{jl}$, i.e., $f\left(\boldsymbol{0}\mid\boldsymbol{\theta}_{jl}\right) = \sqrt{u\left|\boldsymbol{L}^{(jl)}\right|/(2\pi)^K}\exp\left\{-u\left(\boldsymbol{0}-\boldsymbol{\theta}_{jl}\right)^\top \boldsymbol{L}^{(jl)}\left(\boldsymbol{0}-\boldsymbol{\theta}_{jl}\right)/2\right\}$. This interpretation follows the hierarchical framework of \cite{zhe2013joint}, where priors are constructed through pseudo-observations. We define $L^{(jl)}_{k,k}=(K-1)/\left(\theta_{k,jl}^2+\epsilon^2\right)$ and  $L^{(jl)}_{k,k^{\prime}}=-1/\sqrt{\left(\theta_{k,jl}^2+\epsilon^2\right)\left(\theta_{k^{\prime},jl}^2+\epsilon^2\right)}$ for $1\leq k\neq k^{\prime}\leq K$ and $1\leq j<l\leq p$, with $\epsilon$ being a small constant. Thus, for each $\boldsymbol{\theta}_{jl}$, the prior involves $\boldsymbol{\theta}_{jl}^\top\boldsymbol{L}^{(jl)}\boldsymbol{\theta}_{jl}=0.5\sum_{k\neq k^{\prime}}\left(  \theta_{k,jl} /\sqrt{\theta_{k,jl}^2+\epsilon^2}-\theta_{k^{\prime},jl} /\sqrt{\theta_{k^{\prime},jl}^2+\epsilon^2}\right)^2$, which is an approximation to the Bayesian counterpart of the sign-based constraint penalty $0.5\sum_{k\neq k^{\prime}}\left[\operatorname{sgn}(\theta_{k,jl})-\operatorname{sgn}(\theta_{k^{\prime},jl})\right]^2$. Combined with the indicator function ${I}\left\{\boldsymbol{\theta}_{jl}=\boldsymbol{\omega}_{jl}\right\}$, the proposed approach encourages the pairwise relationships of predictors across subgroups to have similar signs, leading to networks with a shared pattern of sparsity as well as edges with shared positive or negative relationships. This significantly advances from the previous research based on the group penalty \citep{danaher2014joint,hao2018simultaneous,li2019bayesian,dai2022bayesian}, which utilizes group selection without appropriately accounting for the similarity of the estimated values and fails to sufficiently address the relationships between different networks. This is also more flexible than the strategy based on fused lasso \citep{danaher2014joint,gao2016estimation,zhang2021efficient}, which promotes both shared directions and magnitudes of edges. 
 
 %Similar prior assumptions can also be applied to $\boldsymbol{\beta}$ and $\boldsymbol{\mu}$ when considering the common structures within regression coefficients and predictors' means.	 %yang2015fused,

 \subsection{Posterior inference} 
 	The	complete posterior distribution of the parameters conditionally on the available data is: 
 	\begin{equation}\label{equa:obj}  
 		\begin{aligned}  f\left(\boldsymbol{\beta},\boldsymbol{\mu},\boldsymbol{\Omega},\boldsymbol{\theta},\boldsymbol{\gamma},\boldsymbol{g}\mid \boldsymbol{t},\boldsymbol{\delta},\textbf{X}; \boldsymbol{\tau},\boldsymbol{\pi_{g}}\right)	\propto& f\left(\boldsymbol{t},\boldsymbol{\delta},\textbf{X}\mid \boldsymbol{\beta},\boldsymbol{\mu},\boldsymbol{\Omega} ,\boldsymbol{\tau},\boldsymbol{g}\right)f\left(\boldsymbol{\beta}\right)f\left(\boldsymbol{\mu}\right) f\left(\boldsymbol{\Omega}\mid \boldsymbol{\gamma}\right)f\left(\boldsymbol{\gamma}\right)  \\&\times  f\left(\boldsymbol{\theta}\mid \boldsymbol{\Omega} \right)f\left(\boldsymbol{0}\mid\boldsymbol{\theta}\right) f\left(\boldsymbol{g} \mid \boldsymbol{\pi_{g}}\right){ \prod_{k=1}^K I\left\{\boldsymbol{\Omega}_k \succ 0\right\}}, 
 		\end{aligned} 
 	\end{equation}
where  $f\left(\boldsymbol{0}\mid\boldsymbol{\theta}\right)=\prod_{j=1}^p\prod_{l>j}f\left(\boldsymbol{0}\mid\boldsymbol{\theta}_{jl}\right)$ and { $I\left\{\boldsymbol{\Omega}_k \succ 0\right\}$ ensures that the precision matrix $\boldsymbol{\Omega}_k$  is positive definite.}
We note that since the distribution $f\left(\boldsymbol{\theta}\mid \boldsymbol{\Omega} \right)$ is the indicator function, there is no need to estimate $\boldsymbol{\theta}$ separately. Figure S1 in the supplementary materials gives a graphical representation of the Bayesian model.	
 	  
{Based on (\ref{equa:obj}), we develop a three-step strategy for posterior network construction. In the first step, we employ an Expectation-Maximization (EM) algorithm to compute the Maximum a Posteriori (MAP) estimate of $\boldsymbol{\Psi} = \operatorname{vec}(\boldsymbol{\beta}, \boldsymbol{\tau}, \boldsymbol{\mu}, \boldsymbol{\Omega})$, which enables subsequent posterior inference on the sparsity structures.
Within the EM algorithm, in the E step, the conditional expectations of the latent variables $\rho_{k,i}= I\{g_i=k\}$ and $\gamma_{k,jl}$ are computed using the parameter estimates from the previous M-step, as detailed in equation (S-1) of the supplementary materials. In the M-step, we update $(\boldsymbol{\Psi},\boldsymbol{\pi}_g)$ by maximizing the conditional expectation of the complete-data log-posterior (S-2). This optimization is performed using coordinate descent and ADMM algorithms, ensuring the positive definiteness of each $\boldsymbol{\Omega}_k$ following \cite{danaher2014joint}.
The algorithm iterates until convergence, yielding the MAP estimate $\hat{\boldsymbol{\Psi}}:=\left(\hat{\boldsymbol{\Theta}},\hat{\boldsymbol{\tau}}\right)=\operatorname{vec}(\hat{\boldsymbol{\beta}},\hat{\boldsymbol{\mu}},\hat{\boldsymbol{\Omega}},\hat{\boldsymbol{\tau}})$. Following \cite{lee2015joint}, we then obtain a thresholded estimate $\tilde{\boldsymbol{\Theta}}=\operatorname{vec}(\tilde{\boldsymbol{\beta}},\tilde{\boldsymbol{\mu}},\tilde{\boldsymbol{\Omega}})$.

%\subsubsection{Structure recovery via posterior inclusion probability}\label{sec:pips}
In the second step, the uncertainty of the edge inclusion is quantified using the marginal MAP plug-in posterior inclusion probability (PIP): $\tilde{\gamma}_{k,jl}=\mathbb{P}(\gamma_{k,jl}=1\mid \tilde{\omega}_{k,jl})$. 
Finally, in the third step, the network sparsity pattern is determined by thresholding $\tilde{\gamma}_{k,jl}$, with entries exceeding a prespecified level such as 0.5 identified as edges. Further details are provided in Section S2 and Algorithm 1 of the supplementary materials.
}

Sensitivity analyses on hyperparameters confirm that our approach is robust across a range of plausible settings. The Bayesian Information Criterion (BIC) is adopted to choose optimal values of $v_0$ and $u$, which is defined as $-2 \log f \left(\boldsymbol{t}, \boldsymbol{\delta}, \mathbf{X} \mid \tilde{\boldsymbol{\Psi}}, \hat{\boldsymbol{\pi}}_g \right) + \log n \hat{S}$ where $\tilde{\boldsymbol{\Psi}}$ and $\hat{\boldsymbol{\pi}}_g$  are the final estimates obtained from the proposed algorithm and $\hat{S}$ denotes the number of nonzero elements in the sparse parameter estimates \citep{hao2018simultaneous,ren2022gaussian}. %$-2 \log f \big(\boldsymbol{t}, \boldsymbol{\delta}, \mathbf{X} \mid \hat{\boldsymbol{\beta}}, \hat{\boldsymbol{\mu}}, \tilde{\boldsymbol{\Omega}}; \hat{\boldsymbol{\tau}}, \hat{\boldsymbol{\pi}}_g \big) + \log n \sum_{k=1}^K \left( \hat{s}_k + \hat{d}_{\boldsymbol{\mu}, k} + \hat{d}_{\boldsymbol{\beta}, k} \right),$ where $\hat{\boldsymbol{\beta}}, \hat{\boldsymbol{\mu}}, \tilde{\boldsymbol{\Omega}}, \hat{\boldsymbol{\tau}}$, and $\hat{\boldsymbol{\pi}}_g$ are the final estimates obtained from the proposed algorithm; $\hat{s}_k = \# \{(j,l): \tilde{\omega}_{k,jl} \neq 0, 1 \leq l < j \leq p\}$, $\hat{d}_{\boldsymbol{\mu},k} = \# \{ j: \hat{\mu}_{k,j} \neq 0, 1 \leq j \leq p \}$, and $\hat{d}_{\boldsymbol{\beta},k} = \# \{ j: \hat{\beta}_{k,j} \neq 0, 1 \leq j \leq p \}$. 
     	We refer to Sections S2 and S3 of the supplementary materials for details.

 	\section{Theoretical properties}
 	Denote the vector of unknown true parameters by $\boldsymbol{\Psi}^\star:=\left(\boldsymbol{\Theta}^\star,\boldsymbol{\tau}^\star\right)=\operatorname{vec}\left(\boldsymbol{\beta}^\star,\boldsymbol{\mu}^\star,\boldsymbol{\Omega}^\star,\boldsymbol{\tau}^\star\right)$. For the $k$th subgroup, let $\mathcal{U}_k^\star=\left\{j:\mu_{k,j}^\ast\neq 0\right\}$ and $ \mathcal{V}_k^\star=\left\{j:\beta_{k,j}^\ast\neq 0\right\}$ denote the index sets of nonzero true predictor means and regression coefficients, respectively, and let $\mathcal{W}_k^\star=\left\{(j,l):j\neq l, \omega_{k,jl}^\ast\neq 0\right\}$ denote the set of nonzero off-diagonal elements of the true precision matrix. For a set $\mathcal{S}$, let $\left|\mathcal{S}\right|$ denote its cardinality. Define %the sparsity parameters as 
    $d_{\boldsymbol{\mu}}=\left|\cup_{k=1}^K \mathcal{U}_k^\star\right|,  
 	d_{\boldsymbol{\beta}} =\left|\cup_{k=1}^K \mathcal{V}_k^\star\right|$, and $ 
 	s=\left|\cup_{k=1}^K \mathcal{W}_k^\star\right|$. %Define $\Xi$ as some non-empty convex set of parameters $\boldsymbol{\Psi}$ in $\mathbb{R}^{Kp^{2}+2Kp+2K}$. Denote ${\rho}_{k,i}^{\boldsymbol{\Psi}}$ as the probability that the $i$th subject belongs to the $k$th subgroup given $\boldsymbol{{\Psi}}$. 
     %Furthermore, for a $q$-dimensional vector $\boldsymbol{\nu}$, denote $\left\|\boldsymbol{\nu}\right\|_{\max} = \max_{1 \leq j \leq q} \left|\nu_j\right|$. 
     For a matrix $\boldsymbol{M} \in \mathbb{R}^{q_1 \times q_2}$, %define its minimum, maximum, and max-induced norms as  $\|\boldsymbol{M}\|_{\min} = \min_{i,j} \left|M_{ij}\right|$,  $\|\boldsymbol{M}\|_{\max} = \max_{i,j} \left|M_{ij}\right|$, and  
define $\|\boldsymbol{M}\|_{\infty} = \max_{i=1, \ldots, q_1} \sum_{j=1}^{q_2} \left|M_{ij}\right|$.  %For a square matrix $\boldsymbol{A} \in \mathbb{R}^{p \times p}$, let $\lambda_{\min}(\boldsymbol{A})$ and $\lambda_{\max}(\boldsymbol{A})$ denote its smallest and largest eigenvalues. The following conditions are assumed.

 { In Section S4 of the supplementary materials, we specify the assumed conditions, including log-likelihood regularity, subgroup separability, parameter boundedness, and the relative growth rates of tuning parameters and $K$ alongside the minimal signal strength assumption.
 Similar conditions have been adopted in prior studies \citep{hao2021sparse,hao2018simultaneous, ren2022gaussian,Zhong2023}.}

 	\begin{theorem}\label{theorem:1} 
    		Let $\tilde{s}=   d_{\boldsymbol{{\beta}}} \sqrt{ d_{\boldsymbol{{\mu}}}}+ d_{\boldsymbol{{\beta}}}^2+\sqrt{ p}+\sqrt{ {s+p/K}} $. Assume that $ \tilde{s}\sqrt{ { K^3\log p} }=o(\sqrt{n})$. Then,	
 		
 		\noindent 1. {Under Assumptions 1-5 (supplementary materials)}, the non-asymptotic bound of the estimation error of $\hat{\boldsymbol{{\Psi}}}$ obtained with the proposed EM algorithm is
 		\begin{equation} \label{e:rate}
 			\left\|\hat{\boldsymbol{\Psi}}-\boldsymbol{\Psi}^{*}\right\|_{2} =   O_p\left(\tilde{s}\sqrt{ { K^3\log p}/{n}}\right).
 		\end{equation} 
 		2. {%Assume the lower bound of the minimal signal in the true parameters satisfies $\min\limits_{k=1, \ldots, K}\left\{\min\limits_{(j,l) \in \mathcal{W}_{k}^\star } |\omega_{k,jl}^{*}|,\min\limits_{j\in \mathcal{U}_{k}^\star } |\mu_{k,j}^{*}|,\min\limits_{j \in \mathcal{V}_{k}^\star } |\beta_{k,j}^{*}|\right\}>2  \tilde{s}\sqrt{ { K^3\log p}/{n}}$.
     Under Assumptions 1-6 (supplementary materials), the thresholded estimators  $\tilde{\boldsymbol{\Theta}} = \hat{\boldsymbol{\Theta}}  \cdot I \left\{ |\hat{\boldsymbol{\Theta}} | > \tilde{s}\sqrt{ K^3\log p / n} \right\}$ satisfies that 
        %For the final  estimators $\tilde{\boldsymbol{\Omega}}_{k}$, $\tilde{\boldsymbol{\mu}}_{k}$, and $\tilde{\boldsymbol{\beta}}_{k}$ with  $\tilde{\omega}_{k,jl} =\hat{\omega }_{k,jl} {I}\left\{\left|\hat{\omega}_{k,jl} \right|> \tilde{s}\sqrt{ { K^3\log p}/{n}}\right\}$,    $\tilde{\mu}_{k,j} =\hat{\mu }_{k,j} {I}\left\{\left|\hat{\mu}_{k,j} \right|> \tilde{s}\sqrt{ { K^3\log p}/{n}}\right\}$, and    $\tilde{\beta}_{k,j} =\hat{\beta}_{k,j} {I}\left\{\left|\hat{\beta}_{k,j} \right|> \tilde{s}\sqrt{ { K^3\log p}/{n}}\right\}$,  
        $\left\|\tilde{\boldsymbol{\Omega}}-\boldsymbol{\Omega}^{*}\right\|_{2} =   O_p\left(\tilde{s}\sqrt{ { K^3\log p}/{n}}\right)$ and %achieves selection consistency in the sense that
$$\mathbb{P}\left(\tilde{\gamma}_{k,jl}={\gamma}_{k,jl}\right)  \rightarrow 1,  \quad \mathbb{P}\left(\widehat{\mathcal{W}}_{k}={\mathcal{W}}_{k},\widehat{\mathcal{U}}_{k}=\mathcal{U}_{k}^\star,\widehat{\mathcal{V}}_{k}=\mathcal{V}_{k}^\star\right)  \rightarrow 1$$
 for any $k=$ $1, \ldots,K$ and $1\leq j<l\leq p$,  %$\tilde{\boldsymbol{\Omega}}=\operatorname{vec}\left( \tilde{\boldsymbol{\Omega}}_1,\ldots, \tilde{\boldsymbol{\Omega}}_K \right)$,
  where $\widehat{\mathcal{W}}_{k}=\{(j,l)\mid \tilde{\gamma}_{k,jl}>a\}$ with a threshold $a\in(0,1)$, %denotes the set of the nonzero off-diagonal elements of $\tilde{\boldsymbol{\Omega}}_{k}$, 
and $\widehat{\mathcal{U}}_{k}$ and $\widehat{\mathcal{V}}_{k}$ denote the sets of the nonzero elements of $\tilde{\boldsymbol{\mu}}_{k}$ and $\tilde{\boldsymbol{\beta}}_{k}$, respectively. } 
 	\end{theorem} 
 {In Theorem \ref{theorem:1}, we establish the existence of a local MAP estimator with the desired estimation properties. } If $K$ is fixed, the non-asymptotic bound of the estimation error reduces to
 	$\left\|\tilde{\boldsymbol{\Omega}}-\boldsymbol{\Omega}^{*}\right\|_{2}=O_p\left(\left( d_{\boldsymbol{{\beta}}} \sqrt{  d_{\boldsymbol{{\mu}}}}+  d_{\boldsymbol{{\beta}}}^2+ \sqrt{ { s+p}}\right)\sqrt{  \log p/n}\right)$. Denote by $\mathcal{\boldsymbol{Q}}\equiv\left\{\boldsymbol{\Omega}=\operatorname{vec}\left( \boldsymbol{\Omega}_1,\ldots,\right.\right.$ $\left.\left.\boldsymbol{\Omega}_K \right): \boldsymbol{\Omega}_k\right.$      is positive definite and
	$\left. \max_k\|\boldsymbol{\Omega}_k\|_{\infty} \leq C_{\mathcal{Q}}\right\}$ the class of precision matrices.  Under Assumption 3, the upper bound $C_{\mathcal{Q}}$ of $\mathcal{\boldsymbol{Q}}$ is independent of $(n, p)$.  Provided that $ d_{\boldsymbol{{\beta}}} \sqrt{  d_{\boldsymbol{{\mu}}}}+  d_{\boldsymbol{{\beta}}}^2=O(\sqrt{s+p})$, the convergence rate of the precision matrix estimate $\left\|\tilde{\boldsymbol{\Omega}}-\boldsymbol{\Omega}^{*}\right\|_{2}=O_p(\sqrt{(s+p)\log p/n})$, which achieves the minimax optimal rate for estimating an $s$-sparse precision matrix under the Frobenius norm as shown in \cite{cai2016estimating}. The proof of Theorem \ref{theorem:1} is provided in Section S5 of the supplementary materials.

{ We establish theoretical guarantees for estimation consistency in a challenging setting characterized by double-level heterogeneity, censored outcomes, and a novel similarity prior—a regime not covered by existing theory. Building on this foundation, we prove the consistency of the MAP plug-in PIPs, thereby providing a rigorous basis for network structure recovery and the quantification of Bayesian edge-connection uncertainty within our framework. This distinguishes our theoretical contributions from prior work: for instance, \cite{gan2019bayesian} analyze a convex single-network model, while \cite{hao2018simultaneous} study an unsupervised multi-network mixture without outcome-driven heterogeneity. In contrast, our results simultaneously address non-convexity and supervised, outcome-dependent heterogeneity, offering novel theoretical support for network construction in this complex setting.}

 	\section{Simulation}
 	\label{s:simulation}
 	Simulation studies are conducted to evaluate the performance of the proposed approach under the following settings. (a)  
 	$K=2$ and $3$ and $p = 100$. (b) Consider two settings for the sample size. One is a balanced design where there are 150 subjects in each subgroup, while the other is an imbalanced design with sample sizes being $100$ and $200$ for $K=2$ and $100, 150$, and $200$ for $K=3$. Thus, there are $n=300$ subjects for $K=2$, and $n=450$ for $K=3$ in total. (c) 
 	The predictor observations $\left\{\boldsymbol{x}_i\right\}$ in the $k$th subgroup are simulated from $\mathcal{N}\left(\boldsymbol{0},\boldsymbol{\Omega}_{k}^{-1}\right)$, where $\boldsymbol{\Omega}_{k}$ is generated based on the corresponding network structure. Specifically, following \cite{danaher2014joint}, we simulate each network with ten unconnected subnetworks. To comprehensively examine the performance of the proposed approach under different degrees of similarity across subgroup-specific precision matrices, we consider three specific settings S1-S3, with the number of subnetworks sharing the same sparsity structure in all subgroups being 3, 5, and 7, respectively. For each subnetwork, we consider three types of structure, including the power-law, nearest-neighbor, and Erd\"{o}s-R\'{e}nyi networks. The detailed settings are provided in Section S6 of the supplementary materials.
 	(d) Each logarithmic survival observation $t_i$ is generated as follows. First, the logarithmic survival times of the $k$th subgroup are generated from $ \mathcal{N}\left(\boldsymbol{\beta}_k^{\top}\boldsymbol{x}_{i}, 0.01^{2}\right)$, where five elements of $\boldsymbol{\beta}_k$ are non-zero, and the rest $p-5$ elements are all zero with $\boldsymbol{\beta}_1=\left(2,2,2,2,2,0,\ldots,0\right)^{\top}$,  $\boldsymbol{\beta}_2=-\left(2,2,2,2,2,0,\ldots,0\right)^{\top}$, and $\boldsymbol{\beta}_3= \left(0,0,1,1,1,1,1,0,\ldots,0\right)^{\top}$, respectively. Then, the censoring time is simulated from a Gamma distribution with the shape and scale parameters controlling the censoring rate to be $\sim 20\%$. There are a total of 36 scenarios.

 	{In addition to the proposed approach SBJGM, we compare six alternative approaches: BCGS \citep{dai2022bayesian}, SCAN \citep{hao2018simultaneous}; two “ideal’’ methods, TGemBag and TJGL, which assume true subgroup memberships and apply group estimation of multiple Bayesian graphical models (GemBag, \cite{yang2021gembag}) and joint graphical lasso (JGL, \cite{danaher2014joint}) to estimate subgroup-specific networks; and two two-stage methods, FMGemBag and FMJGL, which first use the finite-mixture AFT model (FM) of \cite{2019AOAS} to estimate subgroup memberships before applying GemBag and JGL. Details are provided in Section S6 of the supplementary materials.}
 	
 	To evaluate the performance of different approaches, we consider the clustering error (CE) for evaluating the heterogeneity identification performance, precision matrix squared error (PME) for evaluating the estimation performance, and true and false positive rates (TPR and FPR) for evaluating the network identification performance. The details are provided in Section S6 of the supplementary materials. Here, similar to other network-based heterogeneity analysis studies \citep{hao2018simultaneous}, we focus on the performance of estimation and identification on $\omega_{k,jl}$'s, which can be more challenging than on $\mu_{k,j}$'s and $\beta_{k,j}$'s.  
 	% {\color{red}(4) P value computed from the log-rank testing (P-logrank) for evaluating the difference of the survival time distributions between the identified subgroups.}
 	
\begin{figure}[!ht]
        \centering \includegraphics[width=1\linewidth]{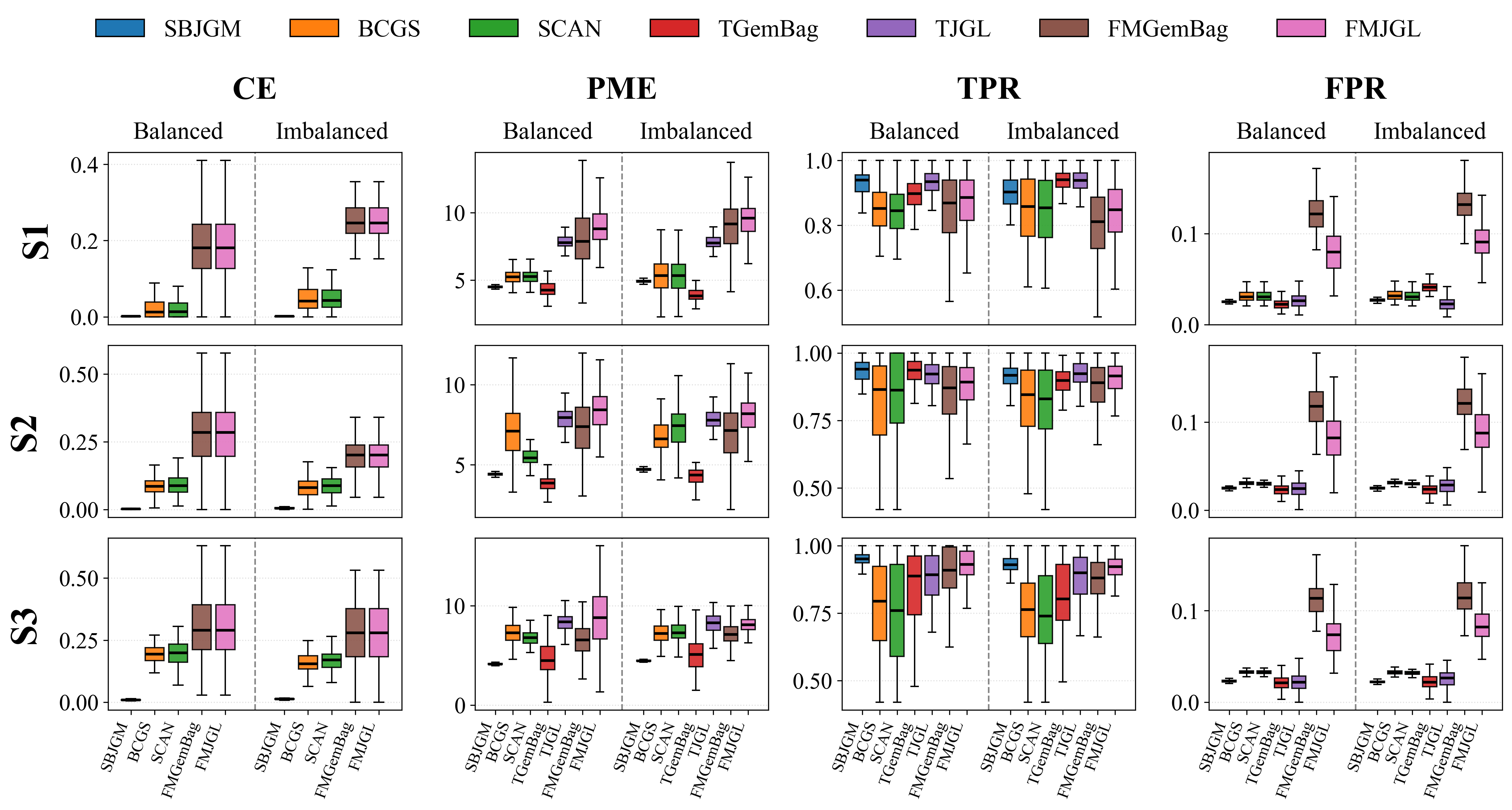}
      	\caption{{Simulation results for the power-law network under the scenarios with $K=2$ based on 100 replicates.}}
    \label{fig: simulation1} 
    \end{figure}

 	100 replicates are conducted for each scenario, where the true value of $K$ is used for all approaches. The summary results for $K=2$ under power-law and Erd\"{o}s-R\'{e}nyi networks are presented in Figures \ref{fig: simulation1} and \ref{fig: simulation2}, and the rest of the results are provided in Section S6 of the  supplementary materials. Under all simulated scenarios, the proposed SBJGM approach consistently demonstrates superior or competitive performance compared to the alternative approaches. In terms of subgroup identification, SBJGM performs significantly better with much smaller CEs. Among the alternatives, the FM methods tend to exhibit the poorest subgroup identification performance, while SCAN and BCGS have better subgroup identification accuracy but are still much inferior compared to SBJGM.

As the similarity between the networks increases (from S1 to S3), it is more challenging to identify subgroup memberships. However, SBJGM still shows more prominent advantages compared to SCAN, BCGS, FMGemBag, and FMJGL in terms of both heterogeneity and network identification performance. It is interesting that, under S3, although with a certain level of clustering errors, SBJGM has better network identification performance than TGemBag and TJGL with ``ideal'' true subgroup memberships, suggesting the effectiveness of the proposed similarity-based prior. Under the scenarios with an imbalanced design or a larger number of subgroups, performance of all approaches declines. Under these scenarios, the proposed approach, when compared to TGemBag, exhibits a slightly worse performance in network identification and estimation due to the more complex unknown heterogeneity and the relatively limited sample size. However, it still consistently outperforms the other alternatives. It maintains its superiority under the scenarios with various types of network structure, including the more challenging power-law networks with denser structures and hub structures. With relatively simpler nearest neighbor networks, when all methods generally perform well with higher TPR values, the proposed approach retains significantly fewer FPs compared to most alternatives. 

 \begin{figure}[!ht]
    \centering
\includegraphics[width=1\linewidth]{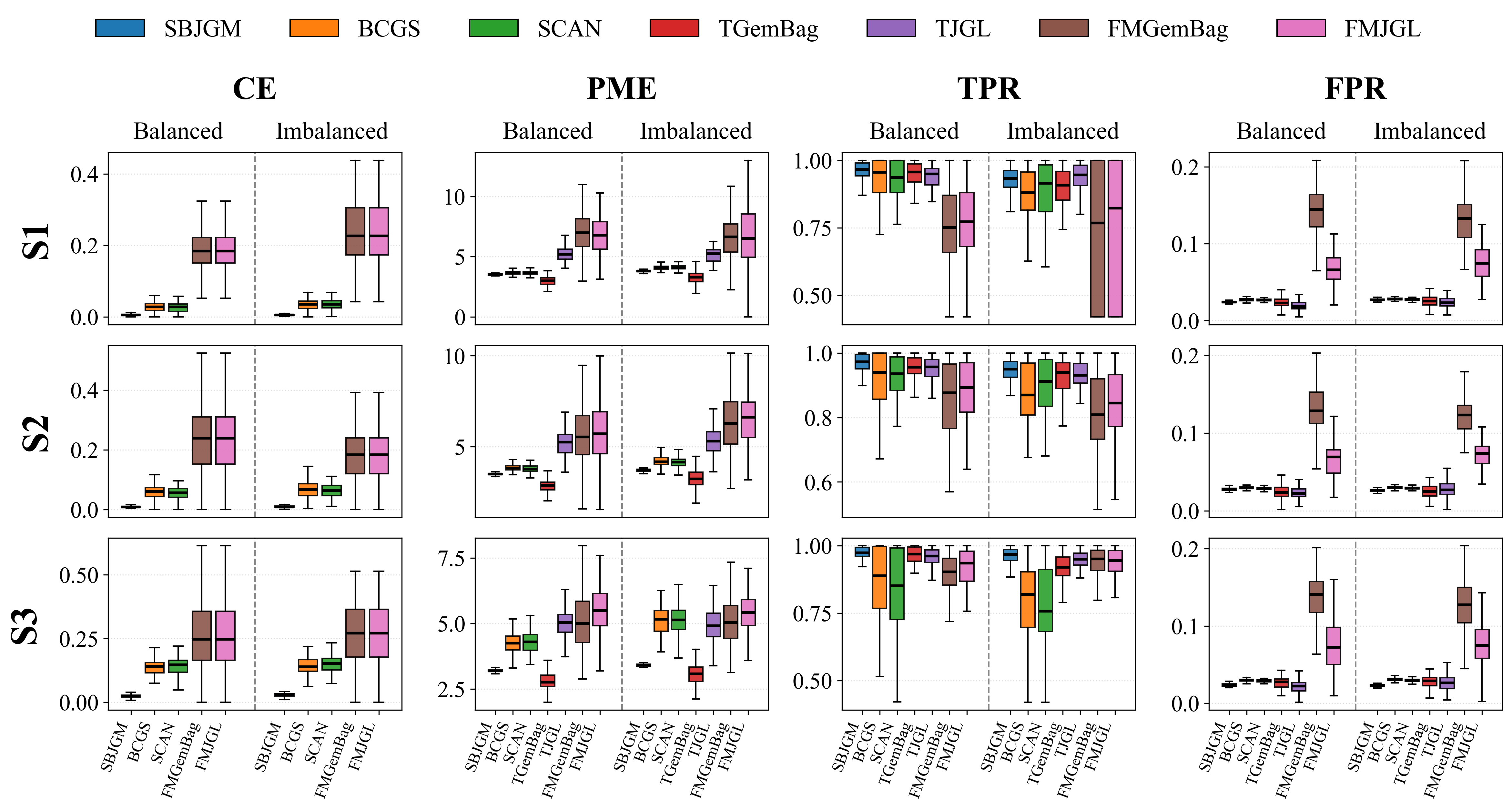}
\caption{{Simulation results for the Erd\"{o}s-R\'{e}nyi network under the scenarios with $K=2$ based on 100 replicates.} }
\label{fig: simulation2}
\end{figure} 

Additionally, to gain a deeper understanding of the similarity-based prior, we illustrate a representative replicate under the scenario with setting S3 and an imbalanced design. The heatmaps depicting the true sparsity structures and the estimated results of the precision matrices with the proposed SBJGM, SCAN, and TGemBag, are presented in Figures S7-S8 (supplementary materials). The findings indicate that SBJGM excels in accurately identifying true sparsity structures and more effectively accommodating the similarity structures across different networks. { To further evaluate uncertainty quantification, Figure S9 displays the PIPs versus the normalized true signals under the same setting. The estimated PIPs effectively capture uncertainty in network sparsity patterns, while a clear separation at the 0.5 threshold confirms the model’s accuracy in detecting true signals.

In addition, we conduct a series of extended simulation experiments, all examined under the power-law and S1 network structures. First, to evaluate performance under an increasing number of subgroups, we consider $K = 4, 5, 6$, and $7$. Second, to investigate the impact of censoring, we examine scenarios with approximately $30\%, 35\%$, $40\%$ and $45\%$ censoring rate under $K = 2$. Third, to assess robustness to error distribution misspecification, we generate survival times from Weibull and log-logistic error distributions with $K=2$. 
The results, presented in Supplementary Figures S10-S12, consistently demonstrate that SBJGM accurately identifies subgroups and recovers network structures across these diverse scenarios.}

Following previous studies \citep{gao2016estimation,hao2018simultaneous,Zhong2023}, the main simulations assume that the number of subgroups $K$ is known for a more fair comparison. When $K$ is unknown, we recommend using the BIC for model selection. As an illustration, we consider scenarios with a power-law network and balanced design and evaluate the empirical performance of BIC over 100 replicates with candidate values $K = 1, \ldots, 5$. The selection frequencies, summarized in Table S2 (supplementary materials), indicate that BIC generally performs well. In particular, the S3 configuration, which exhibits greater subgroup similarity than S1 and S2, poses more challenges for accurate identification. For the BIC-selected values of $K$, Table S3 (supplementary material) reports the corresponding identification and estimation results. In cases where $K$ is incorrectly estimated, label alignment is performed minimizing the estimation error of $\boldsymbol{\Psi}$ \citep{ren2022gaussian}. The results remain consistent with those presented in Figure \ref{fig: simulation1} and Figure S4 (supplementary materials).

 	\section{Data analysis} 
 	\label{s:data analysis}
 	We illustrate the proposed approach in constructing gene networks of different subgroups of cutaneous melanoma (SKCM) from %The Cancer Genome Atlas 
    {TCGA} data. Following existing literature \citep{ahmad2017towards}, we consider the metastatic subjects and use the mRNA gene expression data and overall survival time for the predictors and response, respectively. $360$ subjects are available with $19,039$ gene expression measurements. Among them, $53.6\%$ are non-censored, with the observed survival times ranging from 2.6 to 357.1 months (median 44.05 months). As the sample size is relatively small, we follow the existing studies \citep{hao2018simultaneous} and conduct gene filtering. Specifically, we first identify 2,260 genes with significance levels below $0.01$ from univariate Cox-regression models and then focus on those present in six pathways (pyrimidine metabolism, P53 signaling, cysteine and methionine metabolism, chemokine signaling, cell cycle, and hedgehog signaling pathways) that have demonstrated important implications for SKCM in the literature. With this preprocessing, 91 genes are obtained for downstream analysis. %Data is downloaded from the website https://www.cbioportal.org/. 

 	As the number of subgroups $K$ is unknown, we consider the candidate set $\left\{1,2,3,4,5,6,7\right\}$ and adopt the BIC criterion for selecting the optimal one. With the proposed approach, the BIC criterion leads to two subgroups with sizes of 128 (subgroup 1) and 232 (subgroup 2). The two networks consist of 181 and 345 edges among 80 and 83 genes, respectively. Among them, 52 edges are common and all have the same signs. A graphical representation is provided in Figure \ref{fig:net}, and more detailed estimation results are provided in the supplementary materials. 
  To gain deeper insights into the identified networks, we conduct Gene Ontology (GO) enrichment analysis using DAVID 2021 to examine the functional and biological connections of the related genes. Some significantly enriched GO terms are obtained for the common edges, and genes that exhibit unique edges in the two subgroups are found to be linked to different GO terms. The sensible biological findings provide support to the validity of the proposed network estimation. We refer to Section S7 of the supplementary materials for the detailed GO enrichment analysis results.

	\begin{figure}
		\centering
			\includegraphics[width=1\textwidth]{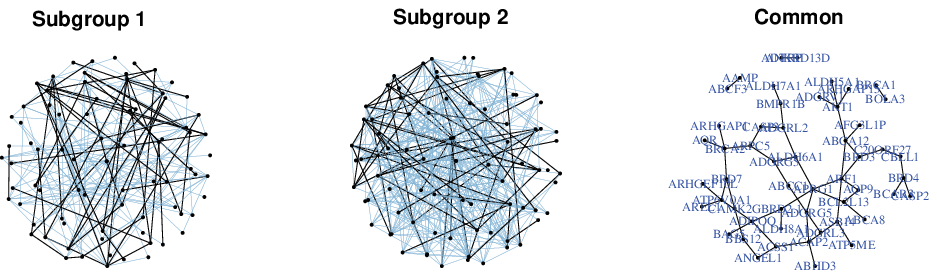} 
 		\caption{Data analysis: gene networks for the two subgroups identified by the proposed approach. In each network, the highlighted edges are shared by the two subgroups.} 
 		\label{fig:net}
	\end{figure}

 	Comparative analysis using the alternative approaches is also carried out with $K=2$. The summary results are reported in Table S4 (supplementary materials), which provide the number of subjects and network edges in the two subgroups identified by different approaches and their overlaps. The subgroups identified using different approaches are matched by correlation. In Table S4 (supplementary materials), different approaches are observed to identify subgroups with different subjects. To get a deeper understanding of the heterogeneity analysis performance, we examine whether these subgroups are associated with survival time. In Figure S13 (supplementary materials), we present the Kaplan-Meier (KM) curves of survival time of different subgroups identified by the proposed and alternative approaches. It is observed that the KM curves of the two subgroups are significantly distinguished from each other for the proposed approach (P value: 0.005). Comparatively, the FM approach shows less significant differences. BCGS and SCAN approaches fail to identify subgroups with distinct survival. These biologically reasonable findings indicate that the proposed approach can satisfactorily identify survival-related subgroups and also provide support for its effectiveness in clinical subgroup identification.

 	In Table S4 (supplementary materials), different approaches demonstrate a moderate number of overlapping edges. To provide indirect support for the network estimation results, we use the resampling strategy and randomly divide the data into a training and a testing set. As in the literature \citep{fan2019assisted}, the values of the negative log-likelihood statistic (NLS) for the 91 genes are calculated to evaluate prediction accuracy. The average NLS values of 100 resamplings are 7102.118 for the proposed approach, 7658.492 for BCGS, 7674.739 for SCAN, 8044.482 for FMGemBag, and 8221.576 for FMJGL, which suggest a satisfactory network estimation accuracy of the proposed approach.

    \section{Discussion}
 	\label{s:discussion} 
 	In this article, we have introduced a novel Bayesian approach for jointly learning multiple sparse networks in a supervised paradigm when the subgroup memberships are unknown. Benefiting from the Bayesian framework, the proposed approach can effectively accommodate both the heterogeneous distributions of predictors and the heterogeneous associations between predictors and survival time, leading to results with stronger clinical implications. In addition to the sparse priors for network estimation, we have introduced a similarity prior to encourage similar network sparsity as well as similar positive or negative effects across different subgroups, which can more effectively address the similarity of subgroups. We have thoroughly examined the theoretical properties of the proposed approach.  A series of numerical experiments have been conducted to demonstrate the superior performance of the proposed approach in both heterogeneity and network identification. By applying the approach to TCGA data, we have discovered various gene networks for SKCM patients and identified biologically relevant relationships associated with clinically sensible subgroups. 

   {In this work, we employ a Bayesian hierarchical framework to accommodate heterogeneity at the molecular level and in molecular-clinical outcome associations. To enhance scalability, we compute MAP estimates via an EM algorithm rather than traditional MCMC. Theoretical guarantees are established based on these MAP estimates, whereas Bayesian uncertainty is quantified through PIPs, which further ensures network recovery consistency. This MAP plug‑in approach is well‑established in prior work, valued for its computational efficiency and solid theoretical foundation \citep{gan2019bayesian, yang2021gembag}. Future directions include exploring MCMC-based approaches for potentially more accurate inference and improving computational efficiency in precision matrix estimation via node-wise strategies \citep{Qin2024multinet}.} We have employed the commonly used Gaussian mixture models for heterogeneity analysis. It can be of interest to explore alternative robust techniques, such as the exponential family graphical models or other nonparametric graphical models, for studying heavy-tailed data. {In this study, we have used the Normal-error AFT model as a representative survival framework. Our proposed approach is general and could be extended to more flexible survival models—such as the Weibull AFT model—by adapting the likelihood accordingly. Furthermore, we have applied Laplace priors on $\beta_{k,j}$ and $\mu_{k,j}$ to maintain a balance between regularization and computational efficiency, since the primary focus here is network construction rather than variable selection among predictors. The spike-and-slab Laplace prior remains a promising alternative for future research that emphasizes predictor-level selection.} Our focus in this data analysis has been mainly on the identification of gene networks using expression data. There is vast potential for a more comprehensive understanding of cancer mechanisms by incorporating other omics measurements, such as mutation and DNA methylation. %proposed by \cite{yang2015graphical} 

% \section{Funding}\label{funding}
% This work was supported by National Natural Science Foundation of China (12071273, 12401382, 12271329, 72331005); Shanghai Rising-Star Program (22QA1403500); Shanghai Science and Technology Development Funds (23JC1402100); Shanghai Research Center for Data Science and Decision Technology; the Program for Innovative Research Team of SUFE; National Institutes of Health (CA204120), and National Science Foundation (2209685).

\section*{Acknowledgments}
The authors thank the editors and reviewers for their invaluable feedback and insightful
suggestions, which have significantly improved this paper. 

% This work was supported by National Natural Science Foundation of China (12071273, 12401382, 12271329, 72331005); MOE Project of Humanities and Social Sciences (25YJCZH291); Shanghai Science and Technology Development Funds (23JC1402100); Shanghai Research Center for Data Science and Decision Technology; the Program for Innovative Research Team of SUFE; National Institutes of Health (CA204120), and National Science Foundation (2209685).

\section*{Disclosure statement}\label{disclosure-statement}

The authors have the following conflicts of interest to declare.

\section*{Data Availability Statement}\label{data-availability-statement}
The data that support the findings in this paper are openly available in TCGA (The Cancer Genome Atlas) at  \href{https://portal.gdc.cancer.gov/projects/TCGA-SKCM}{https://portal.gdc.cancer.gov/projects/TCGA-SKCM}.

\phantomsection\label{supplementary-material}
\bigskip

\begin{center}

{\large\bf SUPPLEMENTARY MATERIAL}

\end{center}
 
				\begin{description}
  \item[Supplement.pdf] Supplement to ``Supervised Bayesian Joint Graphical Model for Simultaneous Network Estimation and Subgroup Identification'', including a graphical illustration of the method, algorithm details, sensitivity results, {assumptions,}  technical proofs,   simulation settings and results, and additional real data results.

  \item[superviseNetcode.zip] The \textsf{R} package \texttt{superviseNet} implementing the proposed method. %, available at \href{https://github.com/mengyunwu2020/superviseNet}{github.com/mengyunwu2020/superviseNet}.
  
  \item[Realdata.xlsx] Network estimation results for real data using the proposed approach. 
\end{description}

\bibliographystyle{chicago}
  \bibliography{bibliography.bib}

\end{document}